\documentclass[aps,prb,twocolumn,showpacs,groupedaddress]{revtex4}
\usepackage{graphicx}

\begin{document}

\title{Quantum bistability and spin current shot noise of a single quantum dot coupled to an optical microcavity}

\author{Ivana Djuric, Marko Zivkovic, Chris P. Search, and Greg Recine }

\address{Department of Physics and Engineering Physics,
Stevens Institute of Technology, Hoboken, NJ 07030}
\begin{abstract}
Here we explore spin dependent quantum transport through a single quantum dot coupled to an optical microcavity.
The spin current is generated by electron tunneling between a single doped reservoir and the dot combined with intradot spin flip transitions induced by a quantized cavity mode. In the limit of strong Coulomb blockade, this model is analogous to the Jaynes-Cummings model in quantum optics and generates a pure spin current in the absence of any charge current. Earlier research has shown that in the classical limit where a large number of such dots interact with the cavity field, the spin current exhibits bistability as a function of the laser amplitude that drives the cavity. We show that in the limit of a single quantum dot this bistability continues to be present in the intracavity photon statistics.  Signatures of the bistable photon statistics manifest themselves in the frequency dependent shot noise of the spin current despite the fact that the quantum mechanical average spin current no longer exhibits bistability. Besides having significance for future quantum dot based optoelectronic devices, our results shed light on the relation between bistability, which is traditionally viewed as a classical effect, and quantum mechanics.
\end{abstract}

\maketitle

\section{Introduction}
Bistability is a phenomenon that readily occurs in classical systems that possess a nonlinear response to some input signal. In a bistable system the output function, $F(I)$, can exhibit two stable states for a certain range of the input $I$ such that when $I$ is varied $F(I)$ follows a hysteresis loop.
One of the most familiar examples is the hysteresis curve in the magnetization of a ferromagnetic material in the presence of an external magnetic field. In the context of electronics, digital flip-flop circuits and Schmitt triggers are common examples of bistable circuits. In nonlinear optics, optical bistability (OB) occurs in the input-output function of an optical resonator that contains a nonlinear dielectric and is driven by a laser \cite{Meystre}. OB has a number of applications in optical communications and computing because it can be used to build all optical switches, logic gates, and optically bistable memory devices\cite{Miller, Abraham2, Gibbs2,Mandel,Waren} but is also interesting for basic studies of phase transitions between stationary but non-equilibrium states \cite{Abraham2,Bonifacio}.

Here we explore a model first proposed by two of us \cite{Djuric-Search-1, IvanaChrisBistable} that unifies research in nonlinear quantum optics with spintronics. In the present work, we use that model to explore how bistability manifests itself in the quantum world. Spintronics has emerged
as a field in which the spin degrees of freedom of charge carriers in solid state devices are exploited for the purpose of
information processing. Manipulation of the spin degrees of freedom rather than the charge has the advantage of longer
coherence and relaxation times since the spin is more weakly coupled to its environment \cite{zutic}. For the same reason, manipulation of the spin of an electron is much harder than the charge and therefore has resulted in significant effort to come up with proposals for necessary spin devices including spin batteries, spin filters, spin transistors, etc... Much of this work has focused on ways to generate pure spin currents, $I_s=s(I_{\uparrow}-I_{\downarrow})$, which are the result of an equal number of spin up ($\uparrow$) and spin down ($\downarrow$) charge
carriers moving in the opposite direction so that the charge current, $I_c=q(I_{\uparrow}+I_{\downarrow})$, is zero. Here,
$I_{\sigma}$ are the spin polarized particle currents, $s=\hbar/2$ the spin of the particle, and $q=e$ the charge. There currently exist
numerous theoretical and experimental concepts for generating spin currents in semiconductor nanostructures including spin-orbit (SO) interactions \cite{extrinsic_SO,rashba}, optical absorption \cite{stevens} and Raman scattering \cite{najmaie}, as well as various types of quantum pumps \cite{mucciolo,watson,sharma,benjamin,blaauboer,sela}.

Electron spin resonance (ESR) between Zeeman states in a quantum dot connected to leads is one of the proposed models for the generation of pure spin currents
\cite{wang-zhang,dong}. According to this model, spin flips are the result of a transverse magnetic field that cause the spin direction of outgoing electrons to be opposite to those entering the dot. Our model \cite{Djuric-Search-1, IvanaChrisBistable} extends this idea to spin flips induced by Raman transitions inside of an optical microcavity. One laser involved in the Raman transition is a strong undepleted pump while the other is a mode of the cavity. Inside of the cavity, both the feedback effect resulting from light "bouncing" back and forth numerous times in the cavity and quantum fluctuations can have a dramatic influence on the characteristics of the spin current. In our previous work \cite{IvanaChrisBistable}, we considered the classical limit of a large number of dots, $N\gg 1$, interacting with the cavity mode such that quantum noise is negligible. When the cavity is driven by a laser, the system exhibits absorptive OB in the amplitude of the cavity field. Because the spin current is a function of the cavity field amplitude, the spin current also exhibits bistability as function of the amplitude of the driving laser which survives even in the presence of significant variations in the dot sizes and coupling to the cavity field.

However, this bistability is a purely classical effect since a large number of dots collectively interact with the cavity mode like a single classical absorber. This begs the question of what happens if we consider only a {\em single} quantum dot coupled to the cavity where quantum fluctuations will be so large as to imply that the two 'stable' outputs lose their stability. Earlier theoretical work in quantum optics that explored the limit of 'bistability' for a single atom
coupled to cavity mode found that the steady state phase space distribution of the cavity field had a bimodal structure indicative of two 'stationary' values \cite{armen,savage,Carmichael,Mabuchi2}. These states are not however stable since quantum noise  forces stochastic jumps between the two values \cite{Carmichael}. Here we show that, while the average spin current for a single dot in a driven cavity does not exhibit bistability, the frequency dependent spin current shot noise does exhibit signatures of the two stationary cavity states since in this system the shot noise spectrum reflects the probability distribution for cavity photon states. In contrast to the mentioned work from quantum optics \cite{armen,savage,Carmichael,Mabuchi2}, which relied on quantum trajectory Monte Carlo simulations, we utilize a standard master equation to calculate the shot noise indicating that evidence of the quantum limit of bistability can be gleaned using more pedestrian techniques.

In Section II, we briefly review our model and introduce our mathematical formulation of the shot noise in terms of the dot+cavity master equation. In Section III, we numerically study both the average spin current and the associated shot noise. In Section IV, we present our conclusions.

\section{Model}
We consider a self-assembled quantum dot embedded in a high-Q microcavity, as depicted in
Fig. 1. We are interested in simultaneous coupling of a dot to a cavity mode and
electrical transport through the dot due to tunneling from a doped reservoir. A number of experiments have already
measured the conductance and shot noise through individual self-assembled quantum dots \cite{schmidt,ota,barthold,kieblich} as well spectroscopy of
exciton and charged exciton states in quantum dots with controllable charging from a doped lead \cite{petroff,atature,warburton}. Other experiments
have demonstrated strong coupling of individual dots to a single optical microcavity mode \cite{reithmaier,peter}. Recently
several of these directions have come together in the experiment by Strauf {\em et al.} \cite{strauf} showing a high efficiency single photon quantum dot source. The experiment demonstrated electrical gate controlled charging of dot, which was embedded in a high-Q optical microcavity, from an n doped layer. Several other experiments have followed demonstrating electrically driven quantum dots embedded in high-Q micropillar cavities that behave as single photon sources \cite{Bockler-APL, Ellis-NewJPhys}.

\begin{figure}[htb]
\includegraphics[height=3.5 in, width=3.5in]{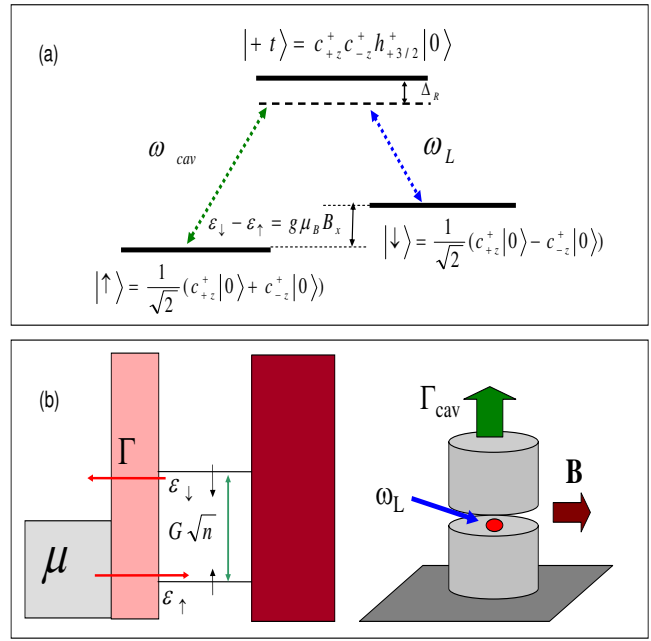}
\caption{(a) Raman transition between the dot Zeeman states,
$|\uparrow,\downarrow \rangle$, via an intermediate trion state, $|+t\rangle$, induced by
a laser with frequency $\omega_L$ and a cavity mode with frequency $\omega_{cav}$. Both optical fields are
detuned from the trion energy by the amount $\Delta_R$ implying that $|+t\rangle$ is a virtual state.
The spin eigenstates along the direction of the magnetic field are superpositions of spin
eigenstates in the growth direction, $\hat{c}^{\dagger}_{\pm z}|0\rangle$. (b) Schematic
of a single quantum dot indicating Zeeman energy levels in the dot and allowed tunneling
between the lead and dot. Also shown to the right is a hypothetical configuration of a dot in a micropillar cavity
showing the direction of the magnetic field, pump laser, and cavity decay.}\label{PIC.1}

\end{figure}

We assume that a single electron reservoir at chemical potential,
$\mu$, is coupled to the dot via tunneling. Only a single empty orbital energy level, $\varepsilon$, of the dot lies close to $\mu$. The Zeeman splitting
between the two electron spin states is
$\Delta=\varepsilon_{\downarrow}-\varepsilon_{\uparrow}=g_x\mu_{B}B$ where $B$ is a
static magnetic field along the x-axis that is perpendicular to the growth direction (z).
$\mu_B$ is the Bohr magneton and $g_x$  is the electronic g-factor along the direction of
the magnetic field. The energy levels satisfy
$\varepsilon_{\uparrow}=\varepsilon-\Delta/2<\mu<\varepsilon_{\downarrow}=\varepsilon+\Delta/2$
so that only spin up electrons can tunnel into the dot and only spin down electrons can
tunnel out. In the limit of very large Coulomb blockade energy, which we consider here,
only a single electron from the reservoir can occupy the dot. The Zeeman states along the direction of the $B$ field are superpositions of spin eigenstates along the growth direction, $|\uparrow,\downarrow\rangle=(1/\sqrt{2})\left( \hat{c}^{\dagger}_{\uparrow_z}|0\rangle \pm \hat{c}^{\dagger}_{\downarrow_z}|0\rangle \right)$, where $\hat{c}^{\dagger}_{\sigma}$ is an electron creation operator.

Raman transitions between the dot Zeeman states, $|\uparrow,\downarrow \rangle$, via an intermediate trion state,
$|+t\rangle$, are induced by a $\sigma^+$ polarized laser with frequency $\omega_L$ and a linearly polarized cavity mode with
frequency $\omega_{cav}$. Several experiments have already demonstrated the use of Raman scattering via an intermediate trion state to manipulate electron spin states in quantum dots \cite{atature,chen,greilich,dutt} and theoretically such processes have been studied inside of optical microcavities for use as a quantum computer \cite{imamoglu}. The $\sigma^{+}$ pump creates a $+3/2$ heavy hole and an electron with spin down along the $z$ direction according to the Hamiltonian, $H_{pump}=(\hbar\Omega_l/2)\exp(-i\omega_lt)\hat{c}_{\downarrow_z}^{\dagger}\hat{h}^{\dagger}_{+3/2}$, which couples to the component of the dot Zeeman states with spin up along $z$ yielding a trion state with an electron singlet. The $\sigma^{+}$ component of the cavity field along with the pump leads to Raman
transitions via the intermediate $|+t\rangle$ state that flips the electron spin while the $\sigma^{-}$ component gives rise to
additional energy shifts due to the AC stark effect. When the two fields are far detuned by an amount $\Delta_R$ from the creation energy for the
$|+t\rangle$ state, the intermediate trion state can be adiabatically eliminated to give
$H_I=\hbar g(\hat{a}\hat{c}^{\dagger}_{\downarrow}\hat{c}_{\uparrow}e^{i\omega_lt}-h.c.)$
where $g=g_{cav}\Omega_l/4\Delta_R$ and $\hat{a}$ is the photon annihilation operator for the cavity mode \cite{Djuric-Search-1,IvanaChrisBistable}.
We have absorbed all energy shifts of the states $|\sigma\rangle$ due to the AC stark effect into a renormalization of the energy levels $\varepsilon_{\sigma}$.
Non-resonant terms $\hat{a}^{\dagger}\hat{c}^{\dagger}_{\downarrow}\hat{c}_{\uparrow}e^{-i\omega_lt}+h.c.$ can be neglected provided $|\Delta- (\omega_{cav}-\omega_l)|\ll |\Delta+(\omega_{cav}-\omega_l)|$.

As one can see in Fig. 1, if an electron enters the dot in the spin $\uparrow$ state, a
photon must be absorbed from the cavity mode and emitted into the pump in order to generate a spin current. It is
therefore necessary to drive the cavity field. We assume that the cavity is driven by a classical source oscillating at frequency $\omega_p$, $H_p=i\hbar \epsilon(\exp(-i\omega_pt)\hat{a}^{\dagger}-h.c.)$, corresponding to coherent coupling between a laser and the cavity mode \cite{Meystre,walls-milburn}.

The Hamiltonian in a frame rotating at the frequency $\omega_p$ is $H'=H'_0+H'_P+H'_I$,
\begin{eqnarray}
H'_0&=&\hbar(\omega_{cav}-\omega_p)\hat{A}^{\dagger}\hat{A}+\varepsilon(\hat{C}^{\dagger}_{\uparrow}\hat{C}_{\uparrow}+\hat{C}^{\dagger}_{\downarrow}\hat{C}_{\downarrow}) \nonumber \\
&+&(\Delta+\omega_l-\omega_p)(\hat{C}^{\dagger}_{\downarrow}\hat{C}_{\downarrow}-\hat{C}^{\dagger}_{\uparrow}\hat{C}_{\uparrow})/2
 \label{H0'}\\
H'_I+H'_P &=& i\hbar g(\hat{A}\hat{C}^{\dagger}_{\downarrow}\hat{C}_{\uparrow}-h.c.)+
i\hbar \epsilon(\hat{A}^{\dagger}-h.c.)
\end{eqnarray}
Here, we have defined operators in a rotating frame $\hat{a}=\hat{A}e^{i\omega_pt}$,
$\hat{c}_{\uparrow}=\hat{C}_{\uparrow}\exp(-i(\omega_l-\omega_p)t/2)$, and
$\hat{c}_{\downarrow}=\hat{C}_{\downarrow}\exp(i(\omega_l-\omega_p)t/2)$.
In this work we assume that the resonance conditions, $\omega_{cav}=\omega_p$ and
$\Delta=\omega_p-\omega_l$, are always satisfied, so that the final Hamiltonian of the
system is $H'=\varepsilon(\hat{C}^{\dagger}_{\uparrow}\hat{C}_{\uparrow}+\hat{C}^{\dagger}_{\downarrow}\hat{C}_{\downarrow})+
i\hbar g(\hat{A}\hat{C}^{\dagger}_{\downarrow}\hat{C}_{\uparrow}-h.c.)+ i\hbar
\epsilon(\hat{A}^{\dagger}-h.c.)$

The dynamics of the system can be described in terms of the density operator, $\rho$, for
the cavity plus dot. The master equation for $\rho$ is given by,
\begin{equation}\dot{\rho}=-i[H',\rho]/\hbar-\Gamma_{cav}(\hat{A}^{\dagger}\hat{A}\rho
-2\hat{A}\rho\hat{A}^{\dagger}+\rho\hat{A}^{\dagger}\hat{A})/2+\dot{\rho}|_{lead}
\label{master}
\end{equation}
The first term describes coherent dynamics of the coupled QD-cavity system, the second
term represents the cavity decay \cite{Meystre,walls-milburn}, and the third term describes
QD-lead coupling. The lead-dot coupling is most easily expressed in terms of the matrix
elements of the density operator, $\rho^{(n,m)}_{\sigma,\sigma'}=\langle
n,\sigma|\rho|\sigma',m\rangle$ where $|\sigma,n\rangle$ represents a state with $n$
photons in the cavity and $\sigma=0,\uparrow,\downarrow$ corresponding to no electrons,
one spin up, or one spin down electron, respectively. The specific form of the master
equations for the lead coupling are \cite{Djuric-Search-1,dong}
\begin{eqnarray}
\dot{\rho}^{(n,m)}_{0,0}|_{lead}&=& \Gamma^{(-)}
\rho^{(n,m)}_{\downarrow,\downarrow}-\Gamma^{(+)}\rho^{(n,m)}_{0,0} \label{lead1}
\\
\dot{\rho}^{(n,m)}_{\uparrow,\uparrow}|_{lead}&=&\Gamma^{(+)}\rho_{0,0}^{(n,m)}
\\
\dot{\rho}^{(n,m)}_{\downarrow,\downarrow}|_{lead}&=&-\Gamma^{(-)}\rho^{(n,m)}_{\downarrow,\downarrow}
\\
\dot{\rho}^{(n,m)}_{\uparrow,\downarrow}|_{lead}&=&-\Gamma^{(-)}\rho^{(n,m)}_{\uparrow,\downarrow}/2.
\label{lead2}
\end{eqnarray}
Here, $\Gamma^{(-)}$ is the rate at which spin down electrons tunnel out of the dot into
lead and $\Gamma^{(+)}$ is the rate at which spin up electrons tunnel into the dot. We
assume that the tunnelling between the lead and the dot is spin independent,
$\Gamma^{(+)}=\Gamma^{(-)}=\Gamma$. We can rewrite Eq. \ref{master} in matrix form,
\begin{equation}
 d \vec{\rho} /dt=M \vec{\rho} \label{rate}
\end{equation}
where $\rho^{(n,m)}_{\sigma_i,\sigma'_i} \rightarrow \vec{\rho}$ is the density matrix in
vector form. The steady state solution, $\vec{\bar{\rho}}$, is given by the eigenvector
of $M$ with zero eigenvalue. Conservation of probability insures that $M$ has a zero eigenvalue \cite{IvanaShotNoise}.

The spin current operator is defined as, $\hat{I}_{s}=s(\hat{I}_{\uparrow}-\hat{I}_{\downarrow})$
with the stationary currents given by $\langle \hat{I}_{\uparrow}\rangle=\Gamma\bar{\rho}_{0,0}$ and $\langle
\hat{I}_{\downarrow}\rangle=-\Gamma\bar{\rho}_{\downarrow,\downarrow}$. Here
$\bar{\rho}_{\sigma,\sigma'}=\sum_n \bar{\rho}^{(n,n)}_{\sigma,\sigma'}$ is the reduced density matrix of the dot after tracing over
the cavity field and $s=\hbar/2$. We note that the spin current can be easily interpreted as the rate at which spin up electrons tunnel into the empty dot, $\Gamma\bar{\rho}_{0,0}$,
plus the rate at which spin down electrons leave the dot, $-\Gamma\bar{\rho}_{\downarrow,\downarrow}$. The average spin current can be expressed in terms of expectation values of the cavity field using Eq. \ref{master},
\begin{equation}
\langle I_s\rangle =2s(2\epsilon Re[\langle \hat{A} \rangle]- \Gamma_{cav}\langle
\hat{A}^{\dagger}\hat{A}\rangle) \label{avspincurrent}
\end{equation}
One sees that the spin current is also the difference between the rate at which photons are coherently injected into the cavity by the driving laser, $2\epsilon Re[\langle \hat{A}\rangle]$, and the rate at which photons decay from the cavity, $\Gamma_{cav}\langle \hat{A}^{\dagger}\hat{A}\rangle$. Conservation of energy, implies that this difference must be absorbed by a spin flip of the electron in the dot.

The noise power spectrum for the current can be expressed as the Fourier transform of the
current-current correlation function,
\begin{equation}
S_{{\sigma},{\sigma'}}(\omega)=2\int_{-\infty}^{\infty}dt e^{i \omega t}[\langle
\hat{I}_{\sigma}(t)\hat{I}_{\sigma'}(0)\rangle - \langle \hat{I}_{\sigma} \rangle \langle \hat{I}_{\sigma'}
\rangle ] \label{ShotNoise}.
\end{equation}
The spin current shot noise, $S^{(s)}=2\int_{-\infty}^{\infty}dt \exp(i \omega t)[\langle
\hat{I}_{s}(t)\hat{I}_{s}(0)\rangle - \langle \hat{I}_{s} \rangle \langle \hat{I}_{s}
\rangle ]$ can be written in terms of the shot noise spectrum for the spin resolved currents as
$S^{(s)}=s^2(S_{\uparrow,\uparrow}+S_{\downarrow,\downarrow}-S_{\uparrow,\downarrow}-S_{\downarrow,\uparrow})$.
It is well known that for currents comprised of uncorrelated particles, the noise power spectrum is Poissonian, $S(\omega)=2q\langle \hat{I}\rangle$, where
$q$ is the quantity transported by each particle in the current $\hat{I}$ \cite{Blanter}, $q=e$ in the case of standard charge currents while in our case the transported quantity is spin $q=s$.  It is often convenient to measure the shot noise relative to the Poissonian noise by defining the Fano factor,
\begin{equation}
F(\omega)=\frac{S^{(s)}(\omega)}{2sI_s}
\end{equation}
where $I_s$ is the average spin current. $F(\omega)>1$ represents super-Poissonian noise while $F(\omega)<1$ represents sub-Poissonian noise.

Here we adopt the numerical method for evaluating Eq. \ref{ShotNoise} developed in Ref. \cite{IvanaShotNoise} for use with master equations of the form Eq. \ref{rate}. Briefly stated, the spectral decomposition of the matrix $M$ is given by $M=\sum_{\lambda} \lambda \hat{P}_{\lambda}$
where $\lambda$ is an eigenvalue of $M$ and $\hat{P}_{\lambda}$ is the projection operator associated with that eigenvalue. This form of $M$ can be used to evaluate the time evolution of the current operators, $\hat{I}_{\sigma}$, and in the end yields the following form for the spin current shot noise.
\begin{equation}
S^{(s)}(\omega)=2sI_s + 2\sum_{\lambda \neq 0} \left( \frac{ Tr[\hat{I}_s \hat{P}_{\lambda} \hat{I}_s\bar{\rho}]}{-i\omega-\lambda} +\frac{Tr[\hat{I}_s \hat{P}_{\lambda} \hat{I}_s\bar{\rho}]}{i\omega-\lambda} \right)
\end{equation}
and the first term, the Poissonian contribution, is calculated from $I_s=Tr[\hat{I}_s\bar{\rho}]$. Here we note that the projection operators can be calculated in terms of the left and right eigenvectors of $M$, $\hat{P}_{\lambda}=\vec{v}_{\lambda} \left(\vec{w}_{\lambda} \right)^{\dagger}$
where $\left(\vec{w}_\lambda \right)^{\dagger} M=\lambda \left( \vec{w}_\lambda \right)^{\dagger}$ define the left eigenvectors while $M\vec{v}_{\lambda}=\lambda\vec{v}_{\lambda}$ define the right eigenvectors. They satisfy the orthonormality relation $\left( \vec{w}_{\lambda_n} \right)^{\dagger} \vec{v}_{\lambda_m}=\delta_{n,m}$. We note that this is a different formulation of the projection operators than appears Ref. \cite{IvanaShotNoise} where the $\hat{P}_{\lambda}=SE_nS^{-1}$, $S$ being the matrix whose columns are the right eigenvectors of $M$ and $E_n$ is a square matrix that has zero entries everywhere except the $(n,n)$ element, which is $1$. However, it is easy to show that these forms are mathematically equivalent.

\section{Results}

We first discuss the behaviour of the intracavity field as a function of the driving field amplitude, $\epsilon$. The intracavity field can be readily visualized
in term of the Q-distribution \cite{Meystre,walls-milburn} for the cavity mode in the steady state as shown in Fig. 2. The Q-distribution is defined
as
\[
Q(\alpha)=\sum_{\sigma=0,\uparrow,\downarrow}\langle\alpha,\sigma|\bar{\rho}|\alpha,\sigma\rangle/\pi
\]
where $|\alpha\rangle$ is a coherent state $\hat{A}|\alpha\rangle=\alpha|\alpha\rangle$. It represents a pseudo-quantum mechanical phase space distribution
for bosonic quantum fields where $Re[\alpha]$ and $Im[\alpha]$, which represent the quadrature components of the field, can be interpreted as the position and momentum, respectively, of a fictitious particle. We note that there exist a number of different pseudo-phase space distributions for bosonic fields whose utility depends on the particular problem \cite{walls-milburn}. We chose the Q-distribution because it is both positive semi-definite and can also be interpreted as a probability distribution, namely the probability of measuring the field in the coherent state $|\alpha \rangle$. This therefore allows qualitative comparisons to classical phase space probability distributions.

In Fig. 2(a), which corresponds to weak driving, there is only single peak around $\alpha\approx 0$. This represents a cavity that is over damped such that all energy injected into the cavity is absorbed by the dot. For larger driving, as in Fig. 2(b) and (c), there are two peaks, one at $\alpha\approx 0$ and another at $Re[\alpha]>0$ and $Im[\alpha]=0$. This represents the bistable situation where the cavity field has two most probable states. By contrast, in Fig. 2(d) one can see that the peak around $\alpha=0$ has completely disappeared and only a peak with $Re[\alpha]>0$ remains when the driving is further increased. This peak corresponds to the case where the cavity driving is so strong that the dot transition is saturated. For a saturated transition, $\bar{\rho}_{0,0}=\bar{\rho}_{\uparrow,\uparrow}=\bar{\rho}_{\downarrow,\downarrow}=1/3$ such that the current obtains the maximum value, $\langle I_S\rangle=s(\Gamma\bar{\rho}_{0,0}+\Gamma\bar{\rho}_{\downarrow,\downarrow})=2s\Gamma/3$. Based on Eq. \ref{avspincurrent}, the approximate location of this second peak in the Q-distribution is then
\begin{equation}
|\alpha|=\frac{(2\epsilon/\Gamma_{cav}) + \sqrt{(2\epsilon/\Gamma_{cav})^2-4\Gamma/3\Gamma_{cav}}}{2} \label{alpha-peak}
\end{equation}
where we note that the last term due to the lead, $4\Gamma/3\Gamma_{cav}$, reduces the cavity field amplitude below the value of an empty cavity (i.e. no absorber in the cavity), $2\epsilon/\Gamma_{cav}$. Fig. 3 shows the peak values of the Q-distribution as a function of $\epsilon$ where one can see that a classic hysteresis loop emerges. This can be compared to the semiclassical solution for the cavity amplitude that ignores quantum fluctuations,
\begin{equation}
|\epsilon|-\Gamma_{cav}|\alpha|/2=\frac{g^2|\alpha|\Gamma}{6g^2|\alpha|^2+\Gamma^2/2} \label{semiclassical-bistability}
\end{equation}
Equation \ref{semiclassical-bistability} is obtained from the equations of motion for the expectation values of the cavity and dot operators by factorizing the expectation values of products of operators such as $\langle \hat{A}^{\dagger}\hat{C}^{\dagger}_{\uparrow}\hat{C}_{\downarrow}\rangle \rightarrow \langle  \hat{A}^{\dagger} \rangle \langle \hat{C}^{\dagger}_{\uparrow}\hat{C}_{\downarrow}\rangle $ (Note that one recovers Eq. \ref{alpha-peak} from Eq. \ref{semiclassical-bistability} in the limit that $g|\alpha| \gg \Gamma$.). One can see in Fig. 3 that in the quantum case, the range of $\epsilon$ values where bistability is present has been shifted to higher values due to quantum fluctuations.

\begin{figure}[htb]
\begin{center}
\includegraphics[height=2.5in,width=3.5in]{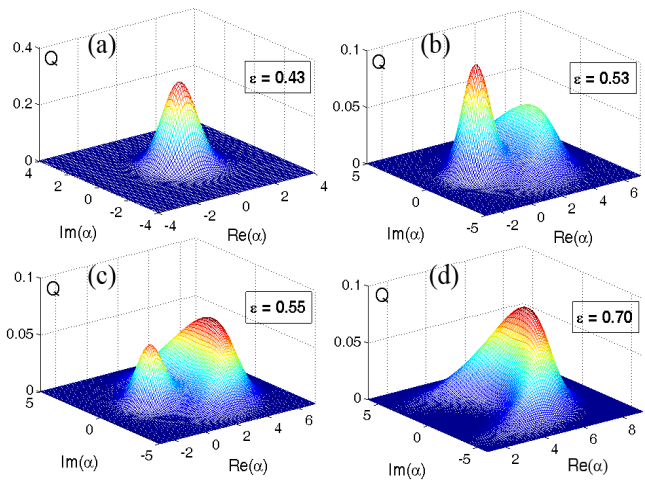}
\caption{(Color Online) Q-distribution vs. $Re[\alpha]$ and
$Im[\alpha]$ for  $\Gamma_{cav}=0.2\Gamma$, $g=2\Gamma$ and in clockwise order (a)
$\epsilon=0.43\Gamma$ (b)$\epsilon=0.53\Gamma$ (c) $\epsilon=0.55\Gamma$, (d) $\epsilon=0.7\Gamma$ }\label{FIG.1a}
\end{center}
\end{figure}

\begin{figure}[htb]
\begin{center}
\includegraphics[height=2.5in,width=3.5in]{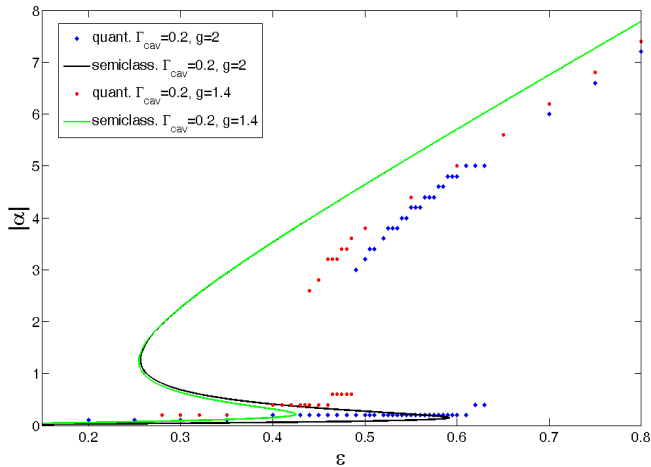}
\caption{(Color Online) Peak values of the Q-distribution as a function of $\epsilon$ (in units of $\Gamma$) for  $\Gamma_{cav}=0.2\Gamma$, $g=1.4\Gamma$ (circles) and $\Gamma_{cav}=0.2 \Gamma$, and $g=2\Gamma$ (diamonds). For comparison, the semiclassical solution Eq. \ref{semiclassical-bistability} exhibiting bistability is also shown for the same parameters (green and black solid lines). One can see that the Q-distribution for a single dot qualitatively follows the semiclassical solution although the range of $\epsilon$ where 'bistability' occurs is reduced by quantum fluctuations. }\label{FIG.1a}
\end{center}
\end{figure}

By contrast, in Fig. 4, we present the quantum mechanical average spin current for a single dot as a function of the driving amplitude. As one can see it is a single valued quantity that shows no sign of the 'switch back' behaviour characteristic of bistability that is seen in the inset, which is the semiclassical spin current calculated using Eq. \ref{semiclassical-bistability} and $I_S=2s(2|\alpha||\epsilon|-\Gamma_{cav}|\alpha|^2)$. In fact, the current is qualitatively the same as that calculated for spin flips in the case of ESR using a classical magnetic field \cite{dong}. This is not surprising since one can see from Eq. \ref{avspincurrent} that the spin current is the quantum mechanical expectation value of the cavity field and despite the bimodal distribution of $Q(\alpha)$, the spin current is averaged over both values, $\langle I_S \rangle\approx P_1 I_{S,1} + P_2 I_{S,2}$ where $P_j$ are the total probabilities corresponding to each of the two peaks in $Q(\alpha)$ and $I_{S,j}=2s(2\epsilon Re[\alpha_j]-
\Gamma_{cav}\alpha^*_j\alpha_j)$ where $\alpha_j$ are the locations of the two peaks.

\begin{figure}[htb]
\begin{center}
\includegraphics[height=2.5in,width=3.5in]{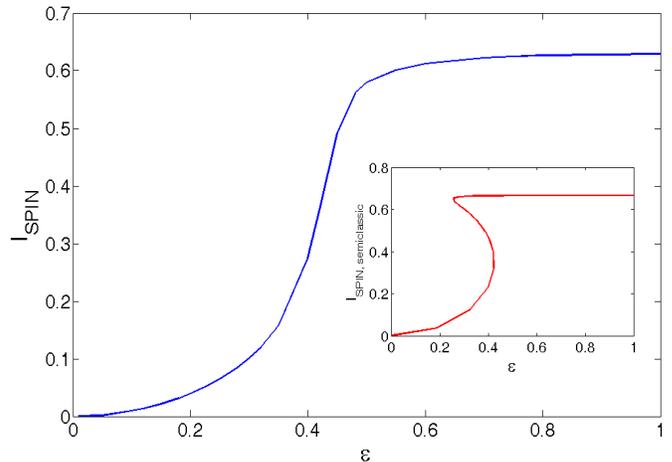}
\caption{(Color Online) Average value of the spin current, $I_s$ (in units of $s\Gamma$) as a function of $\epsilon$ (in units of $\Gamma$) for  $\Gamma_{cav}=0.2\Gamma$, $g=1.4\Gamma$. Inset shows the semiclassical spin current obtained from Eq. \ref{semiclassical-bistability} and $I_S=2s(2|\alpha||\epsilon|-\Gamma_{cav}|\alpha|^2)$. One can see that when quantum fluctuations are included, all indications of bistability are destroyed in the average current.}\label{FIG.1a}
\end{center}
\end{figure}

This begs the question, how does the bistable structure of the intracavity field manifest itself in quantum mechanical observables? Previous work on the quantum limit of bistability for single atom cavity QED focused on the quantum dynamics using 'quantum trajectories' Monte Carlo
simulations approach based on stochastic Schr\"{o}dinger equations \cite{armen} and stochastic master equations
\cite{Carmichael,Mabuchi2}, which showed that the cavity field and photocurrent from the cavity undergo
stochastic jumps between the two states given by the peaks in the Q-distribution. In these systems, the average time between
switching events was proportional to the spontaneous emission lifetime since it was the 'wave function collapse' due to
spontaneous emission of the atom that drove the system between the two states \cite{Carmichael}.

\begin{figure}[htb]
\begin{center}
\includegraphics[height=3.75in,width=3.5in]{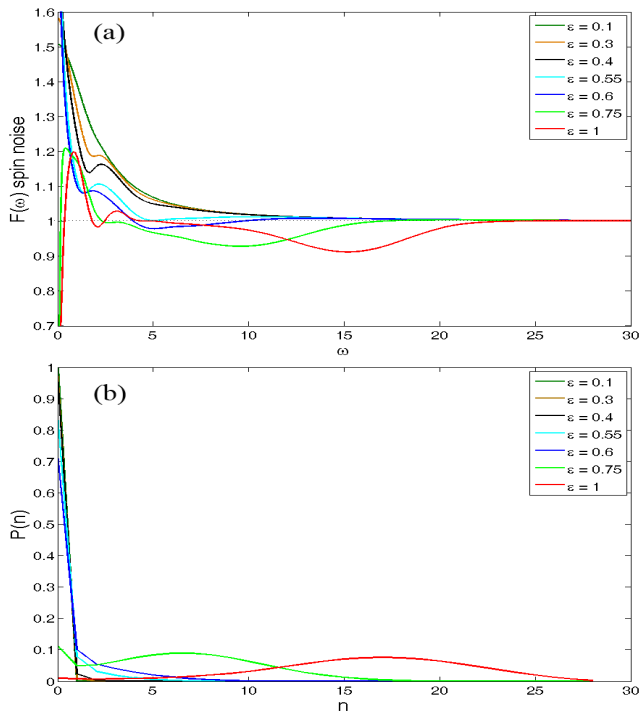}
\caption{(Color Online) (a) Fano factor $F(\omega)$ for $\Gamma_{cav}=0.4\Gamma$, $g=2\Gamma$ and different values of $\epsilon$ (in units of $\Gamma$). (b) Steady state probability distribution for photons in the cavity for the same parameters as in (a). Note that $\omega$ is measured in units of $\Gamma$.}
\end{center}
\end{figure}

\begin{figure}[htb]
\begin{center}
\includegraphics[height=3.75in,width=3.5in]{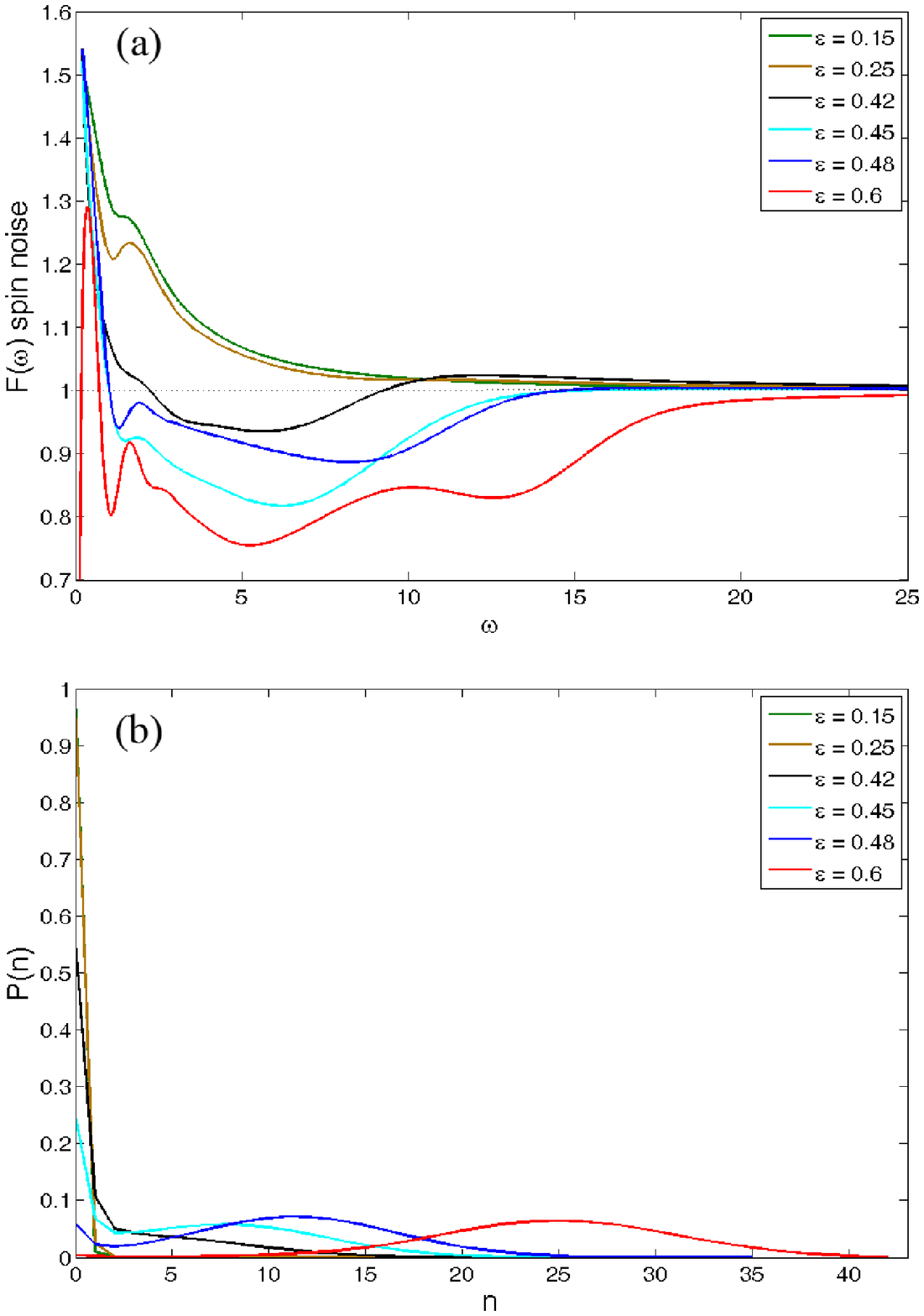}
\caption{(Color Online) (a) Fano factor $F(\omega)$ for $\Gamma_{cav}=0.2\Gamma$, $g=1.4\Gamma$ and different values of $\epsilon$ (in units of $\Gamma$). (b) Steady state probability distribution for photons in the cavity for the same parameters as in (a). Note that $\omega$ is measured in units of $\Gamma$.}
\end{center}
\end{figure}

\begin{figure}[htb]
\begin{center}
\includegraphics[height=2.5in,width=3.5in]{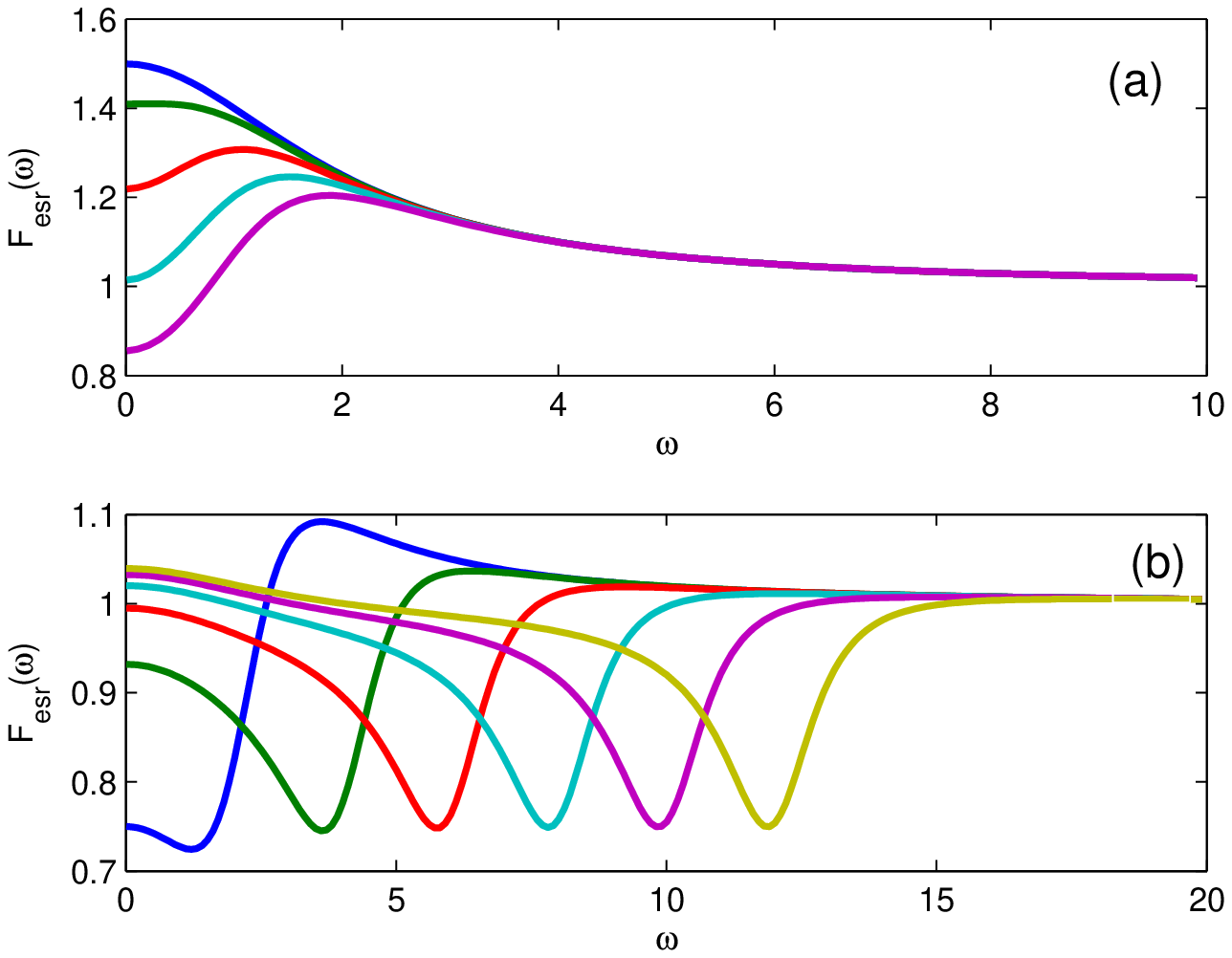}
\caption{(Color Online) Fano factor $F_{esr}(\omega)$ for electron spin resonance with a classical field of Rabi frequency $R$ that flips the spins of the electrons.
The zero frequency Fano factor is given analytically by $F_{esr}(0)=\frac{3\Gamma^4+2\Gamma^2R^2+19R^4}{2(\Gamma^2+3R^2)^2}$. The frequency $\omega$ is in units of $\Gamma$ and for (a) from top to bottom are plotted $R/\Gamma=0$, $R/\Gamma=0.1$, $R/\Gamma=0.2$, $R/\Gamma=0.3$, and $R/\Gamma=0.4$ while for (b) in order of the minima going from left to right $R/\Gamma=1$, $R/\Gamma=2$, $R/\Gamma=3$, $R/\Gamma=4$, $R/\Gamma=5$, and $R/\Gamma=6$. As in the previous graphs, $\omega$ is in units of $\Gamma$.}
\end{center}
\end{figure}

Equations \ref{lead1}-\ref{lead2} have a similar form to that of the master equation for atomic decay. Therefore we can draw an analogy with earlier work and argue
that 'wave function collapse' resulting from electron tunneling events into and out of the dot will induce jumps between the two stable quantum states of the cavity field. Since the time scale that determines transport through the dot is determined by the time needed for a spin flip, which is the Rabi frequency $g\sqrt{n}$,
different cavity field states will result in different time intervals between successive electrons being 'emitted' by the dot. One would therefore expect that the Rabi frequencies associated with the two stable field states would manifest themselves in the current-current correlations, $\langle \hat{I}(t+\tau)\hat{I}(t) \rangle$.

Fig. 5 and Fig. 6 show $F(\omega)$ and $P(n)$, the probability distribution for the cavity photons, for different values of $\epsilon$. For the sake of comparison, Fig. 7 shows the Fano factor for the case of electron spin resonance (ESR), $F_{esr}(\omega)$, with a classical field of Rabi frequency $R$ that flips the spins of the electrons\cite{dong}. The classical field ESR Hamiltonian can be obtained by replacing $\hat{A}$ and $\hat{A}^{\dagger}$ with a c-number ($\hat{A}\rightarrow \alpha$) in $H'$ with $R=g\alpha$. The similarity between Eqs. \ref{lead1}-\ref{lead2} and that of spontaneous emission allows us to define the critical dot number $N_0=2\Gamma\Gamma_{cav}/g^2$, which represents in the semiclassical theory the minimum number of dots necessary for bistability to be present \cite{IvanaChrisBistable,armen}, as well as the critical photon number $n_c=\Gamma^2/4g^2$, which defines the number of photons necessary to significantly modify the dot response \cite{armen}. Classical bistability is predicted to occur in the limit $n_c\rightarrow \infty$ and therefore larger values of $n_c$ should produce more pronounced 'bistability' in the single dot/atom case \cite{savage}. For both Figs. 5 and 6 $N_0\approx0.2$ while for Fig. 5 $n_c=0.063$ and for Fig. 6 $n_c=0.13$. This behavior with $n_c$ is confirmed in the figures where the bimodality of $P(n)$ is more visible and present for a larger range of values in Fig. 6 as compared to Fig. 5.

In these figures, we can see that for small $\epsilon$, below the threshold for the onset of bistability, $F(\omega)$ is super-Poissonian for low frequencies and Poissonian at high frequencies, which is similar to the case of ESR for small $R$ where $F_{esr}(0)\rightarrow 3/2$ for $R\rightarrow 0$. For small $\epsilon$, the cavity is overdamped and only the vacuum state has significant probability, $P(0)$, and therefore transitions are primarily driven by fluctuations above the vacuum state. In the opposite extreme with stronger $\epsilon$ in the bistability region, which is most clearly seen in Fig. 6 for $\epsilon/\Gamma=0.42,0.45,0.48$, $F(\omega)$ remains super-Poissonian at zero frequency while at $\omega\approx 2g|\alpha_2|$ a broad sub-Poissonian dip develops whose overall width is determined by the width of $P(n)$ around the second maximum at $n_2=|\alpha_2|^2$. This behavior is a mixture of the ESR system for small and large R since as already mentioned, $F_{esr}$ is super-Poissonian at low frequencies $R\ll \Gamma$. By contrast, the ESR system exhibits a sub-Poissonian dip at $2R$ for $R>\Gamma$ while being nearly Poissonian at zero frequency ($F_{esr}(0)\rightarrow 19/18$ for $R\rightarrow \infty$). For even larger $\epsilon$ such as $\epsilon/\Gamma=1$ in Fig. 5 or $\epsilon/\Gamma=0.6$ in Fig. 6, which place the system outside of the bistable regime, one can see that the broad sub-Poissonian dip around $\omega\approx 2g|\alpha_2|$ persists but that $F(0)$ is no longer super-Poissonian but rather has become sub-Poissonian. Therefore we can conclude that the super-Poissonian behavior of $F(0)$ is attributable to the maximum in $P(n)$ at $n=0$ while the sub-Poissonian dip is attributable to the the maximum in $P(n)$ around $|\alpha_2|^2$.


\section{Conclusions}
Here we have analyzed the spin current and shot noise from a single quantum dot embedded inside of a driven optical microcavity. We have shown that as a result of the cavity field induced spin flips, the quantum bistability present in the cavity field Q-distribution manifests itself also in the spin current shot noise from the dot. These results indicate that despite the large quantum fluctuations that wipe out all trace of the bistability in the average current, the shot noise reveals the underlying bimodal distribution of the cavity field. This works shows that there is no need to make recourse to more complicated methods such as stochastic wave function methods in order to detect bistability in the presence of large quantum fluctuations.

This work is supported by National Science Foundation.


\end{document}